# Boundary scattering of phonons: specularity of a randomly rough surface in the small perturbation limit

A. A. Maznev[*]

*Department of Chemistry, Massachusetts Institute of Technology, Cambridge MA 02139, USA*

Scattering of normally incident longitudinal and transverse acoustic waves by a randomly rough surface of an elastically isotropic solid is analyzed within the small perturbation approach. In the limiting case of a large correlation length $L$ compared with the acoustic wavelength, the specularity reduction is given by $4\eta^2 k^2$, where $\eta$ is the RMS roughness and $k$ is the acoustic wavevector, which is in agreement with the well-known Kirchhoff approximation result often referred to as Ziman's equation [J. M. Ziman, *Electrons and Phonons* (Clarendon Press, Oxford, 1960)]. In the opposite limiting case of a small correlation length, the specularity reduction is found to be proportional to $\eta^2 k^4 L^2$, with the fourth power dependence on frequency as in Rayleigh scattering. Numerical calculations for a Gaussian autocorrelation function of surface roughness connect these limiting cases and reveal a maximum of diffuse scattering at an intermediate value of $L$. This maximum becomes increasingly pronounced for the incident longitudinal wave as the Poisson's ratio of the medium approaches 1/2 as a result of increased scattering into transverse and Rayleigh surface waves. The results indicate that thermal transport models using Ziman's formula are likely to overestimate the heat flux dissipation due to boundary scattering, whereas modeling interface roughness as atomic disorder is likely to underestimate scattering.

PACS numbers: 43.35.+d, 62.30.+d, 66.70.-f

[*] alexei.maznev@gmail.com

# I. INTRODUCTION

Boundary scattering of phonons has a profound effect on thermal transport in nanostructures [1]. In the simplest model of a perfectly diffuse surface proposed by Casimir [2], a phonon totally "forgets" where it came from and gets scattered with equal probability into any direction. However, any surface tends to become specular for long wavelengths or at grazing incidence angles. The importance of surface specularity was realized early on in studies of thermal conductivity of single crystal rods at low temperatures [3,4]. Subsequently, the specularity parameter, i.e., the probability for a phonon to undergo a specular reflection rather than get diffusely scattered by the surface, became ubiquitous in the analysis of boundary-limited thermal transport [5-7]. More recently, surface specularity at sub-THz frequencies has been studied directly with laser-generated coherent phonons [8,9]. Despite extensive literature on wave scattering from rough surfaces [10-12], a comprehensive analysis of phonon scattering by a randomly rough surface appears to be still lacking. Many researches [13-20] rely on an analytical equation, often ascribed to Ziman [5,20] albeit known earlier [21], that relates the specularity parameter $p$ to the RMS roughness $\eta$, phonon wavevector $k$, and the angle of incidence $\theta$,

$$p = \exp\left(-4\eta^2 k^2 \cos^2\theta\right). \tag{1}$$

Equation (1) reduces the hard problem of wave scattering from a rough surface to a very simple result [22], which, conveniently, does not contain the correlation length of surface roughness $L$. Moreover, it is surmised [5,23] that Eq. (1) is valid for any $L$ as far as specular reflection probability is concerned, with the correlation length only affecting the angular distribution of diffusely scattered phonons. However, in the theory of wave scattering from rough surfaces [8-10] it is well established that Eq. (1) is only valid in the Kirchhoff approximation which assumes that the correlation length is much greater than the wavelength, $kL\gg1$. Indeed, in the opposite limiting case of deeply subwavelength scatterers, $kL\ll1$, one would expect the probability of diffuse scattering to scale as $k^4$ similarly to Rayleigh scattering, in contrast to the $k^2$ dependence according to Eq. (1).

In recent years, a number of advanced and sophisticated models of boundary scattering have been applied to the analysis of thermal transport in nanostructures [25-35]. Many of these studies involve detailed models of a rough surface or interface at the atomic level and use either lattice dynamics calculations based on Green's functions analysis [25-29] or molecular dynamics simulations [32-35]. These advanced studies heavily rely on numerical computations; hence it is difficult to generalize their results beyond specific systems considered in each particular paper. Despite recent advances in atomistic-level modeling, the question of the specularity of a rough surface as a function of the roughness height and correlation length still remains open, and researchers not possessing a sophisticated modeling apparatus still have no tools beyond Eq. (1) at their disposal.

This report aims to address the issue of the surface specularity for a weakly rough surface, i.e. within the small perturbation approach. The latter assumes that the height of surface roughness is small compared to the wavelength, and that the slopes of the surface are small [12], but puts no restrictions on $kL$. In the perturbation approach, the reduction of specularity from unity is assumed to be small, which limits its practical applicability; however it allows one to make progress in the analytical analysis and helps in understanding the main trends, which oftentimes hold even beyond the domain of applicability of the small perturbation approximation.

The perturbation approach has been extensively used to study scattering of scalar waves (such as sound waves in liquid) and electromagnetic waves [10-12,22] as well as to scattering of elastic waves from surfaces with a known profile [36] and attenuation of Rayleigh surface waves on a randomly rough surface [37]. However, very little has been done for the case of elastic wave reflection from a randomly rough surface [12]. A recent study [27] presented perturbation analysis of acoustic wave scattering at rough solid-solid interfaces, but numerical results presented therein hardly allow to draw conclusions beyond the specific cases considered in the study, and the role of the correlation length remained unexplored. Here, we consider the simplest case of a normal incidence of a longitudinal or transverse wave on a weakly rough surface of an elastically isotropic solid which allows us to elucidate general trends and obtain analytical results in limiting cases. We start with a detailed analysis for a longitudinal incident wave, which is compared to the case of a longitudinal wave in liquid, and then extend the analysis to incorporate transverse waves and provide a discussion of oblique incidence.

## II. FORMULATION OF THE PROBLEM

The geometry of the problem is shown in Fig. 1. In the case of a smooth surface the elastic medium occupies the half-space z>0. A normally incident longitudinal wave reflecting from the flat surface z=0 results in a displacement field given by

$$u_z^{(0)} = \frac{1}{\omega}\sqrt{\frac{2}{\rho c_l}}\left(e^{i\omega t + ik_l z} + e^{i\omega t - ik_l z}\right), \qquad (2)$$

where $\omega$ is the angular frequency, $c_l$ is the longitudinal speed of sound, and $k_l = \omega/c_l$ is the wavevector. The first term in parentheses corresponds to the incident and the second term to the reflected wave. The amplitude factor here is chosen to make the incident acoustic power per unit surface area equal to unity.

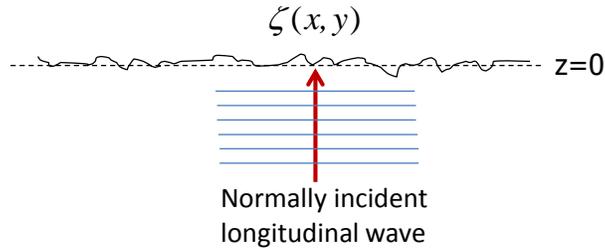

FIG. 1. (Color online) Geometry of the problem.

Let us now consider a rough surface described by a surface profile $\zeta(x,y) = \zeta(\mathbf{r})$ describing a small deviation from z=0. The perturbation approximation [12] requires that $k_l \zeta \ll 1$ and $\nabla \zeta \ll 1$. To simplify subsequent calculations, we assume that roughness occupies a unit area, the surface being flat outside this area. The Fourier transform (FT) of $\zeta(\mathbf{r})$ is given by

$$\tilde{\zeta}(\mathbf{k}) = \int \zeta(\mathbf{r}) e^{i\mathbf{k}\mathbf{r}} d\mathbf{r}. \qquad (3)$$

The RMS roughness η is given by

$$\eta^2 = \overline{\zeta^2(\mathbf{r})} = \frac{1}{4\pi^2}\int \tilde{\zeta}^*(\mathbf{k})\tilde{\zeta}(\mathbf{k})\, d\mathbf{k}, \qquad (4)$$

where * stands for complex conjugate. We introduce a normalized autocorrelation function

$$C(\mathbf{r}_1) = \frac{1}{\eta^2}\overline{\zeta(\mathbf{r})\zeta(\mathbf{r}-\mathbf{r}_1)}, \qquad (5)$$

whose FT is given by

$$\tilde{C}(\mathbf{k}) = \frac{1}{\eta^2}\tilde{\zeta}^*(\mathbf{k})\tilde{\zeta}(\mathbf{k}). \qquad (6)$$

A "well-behaved" autocorrelation function $C(\mathbf{r})$ is characterized by a correlation length $L$ such that $C(\mathbf{r})$ is significantly nonzero at $r \leq L$ and vanishes at $r \gg L$. The spectral autocorrelation function $\tilde{C}(\mathbf{k})$ is significantly nonzero at $q \leq 1/L$ and vanishes at $q \gg 1/L$. In numerical examples below we will be using a Gaussian autocorrelation function,

$$C(\mathbf{r}) = e^{-r^2/L^2}, \quad \tilde{C}(\mathbf{k}) = \pi L^2 e^{-k^2 L^2/4}. \tag{7}$$

The surface roughness results in diffusely scattered waves (i.e. waves, propagating in other directions than surface normal) as well as in a reduction in the amplitude (and possibly, a phase shift) of the specularly reflected wave. Our goal is finding the specularity parameter, equal to the power of the specularly reflected wave per unit area (considering that the power of the incident wave is unity). However, finding the diffusely scattered waves in the first-order perturbation approximation is easier than finding a correction to the specularly reflected field which corresponds to the 2$^{nd}$ order in the perturbation [10]. Therefore we adopt the following approach: we will find the total power of diffusely scattered waves $f$, which has been referred to as the roughness parameter [4], i.e. the probability that an incident phonon is scattered diffusely. The specularity parameter is then found as $p = 1 - f$.

### III. PERTURBATION ANALYSIS FOR LONGITUDINAL WAVE

We represent the displacement field as the sum of the zeroth-order solution given by Eq. (2) and the scattered field whose amplitude is proportional to the amplitude of the surface roughness, $\mathbf{u} = \mathbf{u}^{(0)} + \mathbf{u}^{(1)}$. The boundary conditions require that normal and tangential stress components at the free surface $z=\zeta(\mathbf{r})$ vanish. We follow Gilbert and Knopoff [36] by expanding stresses in a Taylor series at $z=0$ and retaining only terms of the first order in the perturbation, which leads to the following boundary conditions for stress components at $z=0$,

$$\sigma_{zz}^{(1)} = -\zeta \frac{\partial \sigma_{zz}^{(0)}}{\partial z},$$

$$\sigma_{xz}^{(1)} = -\zeta \frac{\partial \sigma_{xz}^{(0)}}{\partial z} - \frac{\partial \zeta}{\partial x} \sigma_{xx}^{(0)} - \frac{\partial \zeta}{\partial y} \sigma_{xy}^{(0)}, \tag{8}$$

$$\sigma_{yz}^{(1)} = -\zeta \frac{\partial \sigma_{yz}^{(0)}}{\partial z} - \frac{\partial \zeta}{\partial y} \sigma_{yy}^{(0)} - \frac{\partial \zeta}{\partial x} \sigma_{xy}^{(0)}.$$

For the zero-order solution given by Eq. (2), shear stresses $\sigma_{xz}^{(0)}$ and $\sigma_{xy}^{(0)}$ are identically zero, and all stress components $\sigma_{ij}^{(0)}$ are zero at $z=0$, which eliminates the right-hand sides in two bottom lines of Eq. (8), leading to the following boundary condition at $z=0$,

$$\sigma_{zz}^{(1)} = -\zeta \frac{\partial \sigma_{zz}^{(0)}}{\partial z},$$
$$\sigma_{xz}^{(1)} = \sigma_{yz}^{(1)} = 0. \tag{9}$$

From Eq. (2), we find

$$\sigma_{zz}^{(0)} = i\sqrt{2\rho c_l}\left(e^{i\omega t + ik_l z} - e^{i\omega t - ik_l z}\right), \tag{10}$$

which leads to

$$\sigma_{zz}^{(1)}\big|_{z=0} = 2^{3/2} k_l \sqrt{\rho c_l}\, \zeta(\mathbf{r}) e^{i\omega t}. \tag{11}$$

Thus the problem of finding the scattered field is reduced to finding waves produced by a harmonic vertical force acting on the flat surface. In order to find the total power of scattered waves we only need to find the displacement field at $z=0$. The spatial Fourier transform of the surface displacement can be expressed in terms of the spectral surface Green's function as follows,

$$\tilde{u}_z^{(1)}\big|_{z=0} = -2^{3/2} k_l \sqrt{\rho c_l}\, \tilde{\zeta}(\mathbf{k})\tilde{G}_{33}(\mathbf{k},\omega) e^{i\omega t}, \tag{12}$$

where $\tilde{G}_{33}(\mathbf{k},\omega)$ is the Fourier transform of the surface Green's function $\tilde{G}_{33}(\mathbf{r},t)$ expressing the vertical surface displacement response to an instantaneous vertical point force acting on the surface. For an elastically isotropic half-space, spectral surface Green's functions $\tilde{G}_{ij}(\mathbf{k},\omega)$ have been obtained in closed form [37,39].

The total power $f$ radiated into scattered waves is given by the product of the effective force acting on the surface and the surface velocity $-\sigma_{zz}^{(1)}\left(\partial u_z^{(1)}/\partial t\right)$, taken at $z=0$, averaged over $t$ and integrated over $\mathbf{r}$,

$$f = \frac{1}{4\pi^4} k_l^3 \rho c_l^2 \operatorname{Re} \iiint i\tilde{G}_{33}(\mathbf{k}_1,\omega)\tilde{\zeta}(\mathbf{k}_1)\tilde{\zeta}^*(\mathbf{k}_2) e^{-i\mathbf{k}_1\mathbf{r}} e^{i\mathbf{k}_2\mathbf{r}} d\mathbf{k}_1 d\mathbf{k}_2 d\mathbf{r}. \tag{13}$$

Integrating over $\mathbf{r}$ yields a delta-function $\delta(\mathbf{k}_1-\mathbf{k}_2)$, which leads to the following result

$$f = \frac{1}{\pi^2} \eta^2 k_l^3 \rho c_l^2 \int \tilde{C}(\mathbf{k}) \operatorname{Im}\tilde{G}_{33}(\mathbf{k},\omega) d\mathbf{k}. \tag{14}$$

Here we have assumed that the autocorrelation function possesses an inversion symmetry (a natural assumption for random roughness), therefore its Fourier-transform is a real function. Thus the specularity parameter $p = 1 - f$ can be found from the spectral autocorrelation function $\tilde{C}(\mathbf{k})$. Note that Eq. (14) is equally applicable to random surfaces and surfaces of known shape. For example, a single Gaussian bump will yield the same scattered power as a randomly rough surface with the same Gaussian autocorrelation function.

For an isotropic half-space, the spectral surface Green's function is given by [37,39]

$$\tilde{G}_{33} = \frac{1}{\rho c_t^2} \frac{k_t^2 \left(k^2 - k_l^2\right)^{1/2}}{R(k)} + \frac{i\pi F}{\rho c_t^2} \delta(\mathrm{k} - k_R), \tag{15}$$

where

$$R(k) = 4k^2 \left(k^2 - k_l^2\right)^{1/2} \left(k^2 - k_t^2\right)^{1/2} - \left(2k^2 - k_t^2\right)^2, \tag{16}$$

$k_l = \omega / c_l$, $k_t = \omega / c_t$, and $k_R = \omega / c_R$ are the longitudinal, transverse and Rayleigh wave vectors, respectively, with the Rayleigh surface velocity $c_R$ found from the Rayleigh equation $R(k_R) = 0$, and $F$ is a dimensionless parameter given by

$$F = \frac{\beta^2}{2} \left(1 - \alpha^2\right)^{1/2} \left[\frac{8(2 - \alpha^2 - \beta^2)}{(2 - \beta^2)^2} + \beta^4 - 4\right]^{-1}, \tag{17}$$

where $\alpha = c_R / c_l$, $\beta = c_R / c_t$. The imaginary branches of square roots in Eqs. (15) and (16) are defined by

$$\left(k^2 - k_{l,t}^2\right)^{1/2} \equiv -i\left(k_{l,t}^2 - k^2\right)^{1/2}, \text{ if } k_{l,t} > k. \tag{18}$$

The delta-function contribution at the pole $k=k_R$ has been added to ensure the causality of the Green's function [40]. Plugging Eq. (15) into Eq. (14) and assuming an isotropic autocorrelation function, we obtain the final result,

$$f = \frac{2}{\pi s^3} \eta^2 k_l^4 \left[ \int_0^s \tilde{C}(xk_t) \frac{x\left(s^2 - x^2\right)^{1/2}}{4x^2 \left(1 - x^2\right)^{1/2} \left(s^2 - x^2\right)^{1/2} + \left(2x^2 - 1\right)^2} dx \right.$$
$$\left. + \int_s^1 \tilde{C}(xk_t) \frac{4x^3 \left(1 - x^2\right)^{1/2} \left(x^2 - s^2\right)}{16x^4 \left(1 - x^2\right)\left(x^2 - s^2\right) + \left(2x^2 - 1\right)^4} dx + \frac{\pi}{\beta} F\tilde{C}(k_R) \right]. \tag{19}$$

where $s = c_t / c_l$. The first two terms in brackets represent the power scattered into bulk waves (with the second term involving transverse waves only), whereas the third term yields the contribution of Rayleigh surface waves.

### A. Limiting cases

Let us consider limiting cases of large and small correlation lengths. To analyze the case of a large correlation length compared to the acoustic wavelength, $k_l L \gg 1$, it is convenient to return to Eq. (14). Since $\tilde{C}(\mathbf{k})$ is only nonzero at very small wavevectors compared to $k_{l,t}$, we can replace $\tilde{G}_{33}(\mathbf{k}, \omega)$ by its value at $\mathbf{k}=0$,

$$\tilde{G}_{33}(\mathbf{k} = 0, \omega) = \frac{i}{\rho c_l \omega} , \qquad (20)$$

which leads to the following result,

$$f_\infty = \frac{1}{\pi^2} \eta^2 k_l^2 \int \tilde{C}(\mathbf{k}) \, d\mathbf{k} , \qquad (21)$$

with the subscript "∞" indicating the infinite correlation length limit. According to the definition of the autocorrelation function, $\int \tilde{C}(\mathbf{k}) \, d\mathbf{k} = 4\pi^2$, hence we obtain

$$f_\infty = 4\eta^2 k_l^2 , \qquad (22)$$

yielding a specularity parameter $p = 1 - 4\eta^2 k_l^2$, which perfectly agrees with Eq. (1). Thus in the limit of a large correlation length the perturbation approach agrees with the Kirchhoff approximation result, as has already been demonstrated for scalar waves and electromagnetic waves [12,22,41].

In the opposite limiting case of a small correlation length, $k_l L, k_t L \ll 1$, we can replace $\tilde{C}(\mathbf{k})$ by $\tilde{C}(\mathbf{k} = 0)$, with the following result,

$$f_0 = \frac{2}{\pi s^3} \eta^2 k_l^4 \tilde{C}(k = 0) \left[ I_{bulk} + I_R \right], \qquad (23)$$

where $I_{bulk}$ and $I_R$ are dimensionless constants on the order unity determined by the velocities ratio $s$ (see Fig. 2),

$$I_{bulk} = \int_0^s \frac{x(s^2-x^2)^{1/2}}{4x^2(1-x^2)^{1/2}(s^2-x^2)^{1/2}+(2x^2-1)^2}dx + \int_s^1 \frac{4x^3(1-x^2)^{1/2}(x^2-s^2)}{16x^4(1-x^2)(x^2-s^2)+(2x^2-1)^4}dx,$$

$$I_R = \frac{\pi}{\beta}F.$$
(24)

In the case of a Gaussian autocorrelation given by Eq. (7) we obtain

$$f_0 = \frac{2}{s^3}\eta^2 k_l^4 L^2 (I_{bulk}+I_R).$$
(25)

Since for a well-behaved autocorrelation function $\tilde{C}(k=0)$ is on the order of $L^2$, this result is quite general even though the numerical factor may vary. As expected, for a small correlation length we come to the Rayleigh scattering limit with the scattering power scaling as $k^4$, as opposed to the $k^2$ dependence found in the limit of a large correlation length. Compared to the Kirchhoff approximation limit given by Eq. (22), there is an extra factor of $k_l^2 L^2$. Thus in the small $k_l L$ limit the diffuse scattering probability is much smaller than the Kirchhoff approximation predicts. The relative values of $I_{bulk}$ and $I_R$ indicate relative contributions of bulk and Rayleigh waves to the total scattered power, and we can see from Fig. 2 that the Rayleigh wave contribution is greater than that of bulk waves. We note that in the limit of a zero correlation length the autrocorrelation function turns into the Dirac delta-function, and the spectrum of the scattered waves is equivalent to the well-studied case of the radiation by a vertical point force [42]. In particular, it is known [42,43] that for $s=1/\sqrt{3}$, the fraction of energy radiated into Rayleigh waves amounts to about 67.4%, which is in agreement with our results.

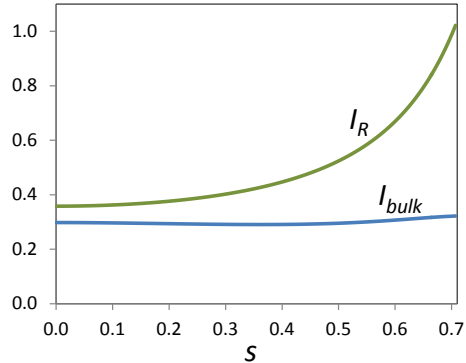

FIG. 2. (Color online) Dependence of dimensionless parameters $I_{bulk}$ and $I_R$ on the transverse-to-longitudinal velocities ratio $s$.

## B. Numerical results for the general case

Let us now consider the general case which requires a numerical evaluation of integrals in Eq. (19). Figure 3 shows the behavior of the diffuse scattering probability $f$ normalized on the infinite correlation length limit $f_\infty$ for the Gaussian autocorrelation function. One might expect the numerical calculations to smoothly connect the limiting cases, with $f$ a monotonically increasing function of $k_l L$. However, the results reveal a maximum at an intermediate value of $k_l L$, which becomes increasingly pronounced at small values of $s$ (for an elastically isotropic medium $s$ can vary between zero and $1/\sqrt{2}$ which corresponds to Poisson's ratio range from 0.5 to 0). This maximum results from scattering into transverse and Rayleigh waves, which is absent in the limit of large $k_l L$ (i.e., in the Kirchhoff approximation). Indeed, if $s$ is small, the wavelengths of transverse and Rayleigh waves are much smaller than the longitudinal wavelength; consequently, even if the roughness height is very small compared to the wavelength of the incident longitudinal wave, it may be not so small compared to the wavelengths of the scattered transverse and Rayleigh waves.

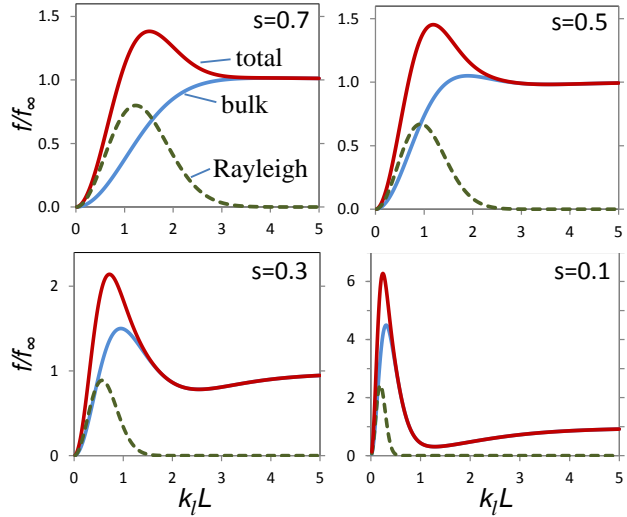

FIG. 3. (Color online) Normalized diffuse scattering probability vs. the product of the acoustic wavevector and the correlation length for different values of the velocities ratio $s$. Contributions of scattering into bulk and Rayleigh surface waves are shown as indicated in the upper left panel.

Even though for typical "hard" solids the velocities ratio $s$ normally exceeds 0.3, there are many examples of soft materials with very low $s$ (such as rubber), for which the maximum of $f$ will occur at small values of $k_l L$ and will greatly exceed the value predicted by Ziman's formula.

In particular, soft soils may have a very low transverse velocity close to the surface [44], hence the issue of increased scattering at small $k_l L$ may be relevant for seismic surveying.

### C. Comparison with the case of a liquid medium

It is instructive to consider, for comparison, the case of a liquid medium in which transverse and Rayleigh waves are absent [45]. The surface Green's function for a liquid half-space is easily obtained from Eq. (15),

$$\tilde{G}_{33} = \frac{1}{\rho\omega^2}\left(k^2 - k_l^2\right)^{1/2}, \qquad (26)$$

which yields a known result [46] for a Gaussian autocorrelation function,

$$f = \eta^2 k_l^4 L^2 \int_0^1 e^{-\frac{k_l^2 L^2 t}{4}} (1-t)^{1/2} dt. \qquad (27)$$

In the limiting case $k_l L \gg 1$ we get the same result as for a solid medium given by Eq. (22), whereas in the limit $k_l L \ll 1$ we get $f_0 = (2/3)\eta^2 k_l^4 L^2$. Numerical calculations for the general case are shown in Fig. 4. In contrast to the case of a solid medium, the maximum of diffuse scattering in the liquid case occurs in the Kirchhoff approximation limit $k_l L \to \infty$. The comparison drives home the point that it is only in the Kirchhoff approximation that the specularity, for a given roughness and acoustic wavelength, is the same for waves of any nature. In the opposite limiting case of a small correlation length, the scaling of $f$ as $\eta^2 k^4 L^2$ is also universal, but the numerical factor depends on the physical system. In fact, for a solid with $s = 0.5$ the numerical factor is almost 20 times larger than for a liquid and will be larger yet for a smaller $s$.

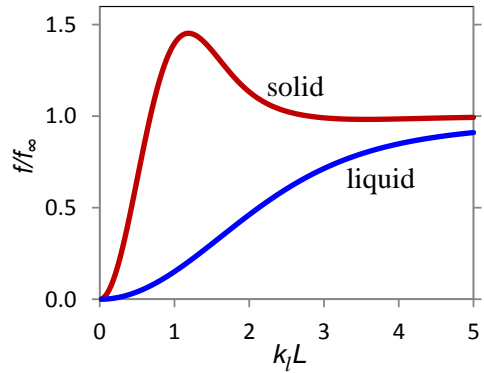

FIG. 4. (Color online) Normalized diffuse scattering probability for a solid with $s=0.5$ vs. a liquid as a function of $k_l L$.

## IV. TRANSVERSE INCIDENT WAVE

The analysis for a normally incident transverse wave parallels the analysis for a longitudinal wave in Sec. III. We consider an incident transverse wave polarized along x, in which case the only non-zero stress component produced by the flat-surface solution is $\sigma_{xz}^{(0)}$. Following the same sequence of steps as in Sec. III, we arrive to the following result for the scattered power,

$$f = \frac{1}{\pi^2}\eta^2 k_t^3 \rho c_t^2 \int \tilde{C}(\mathbf{k}) \operatorname{Im}\tilde{G}_{11}(\mathbf{k},\omega) d\mathbf{k} . \tag{28}$$

which parallels Eq.(14), with the replacement of Green's function $G_{33}$, describing the surface displacement response to a vertical force, by $G_{11}$, which describes the horizontal displacement response to a horizontal force. The final result that parallels Eq. (19) is presented in Appendix.

In the limiting case of a large correlation length $k_t L \gg 1$, we get the familiar Kirchhoff approximation result

$$f_\infty = 4\eta^2 k_t^2 , \tag{29}$$

whereas in the opposite limiting case of a small correlation length $k_t L \ll 1$, we obtain a result that parallels Eq. (23) but has a different numerical factor,

$$f_0 = \frac{1}{\pi}\eta^2 k_t^4 \tilde{C}(k=0)[J_{bulk} + J_R] . \tag{30}$$

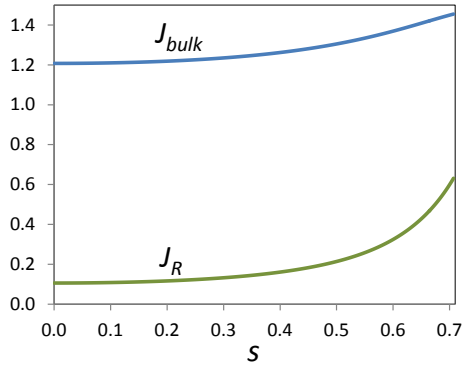

FIG. 5. (Color online) Dependence of dimensionless parameters $J_{bulk}$ and $J_R$ from Eq. (30) on the transverse-to-longitudinal velocities ratio *s*.

The expressions for dimensionless factors $J_{bulk}$ and $J_R$ are presented in Appendix and their dependence on the velocities ratio $s$ is shown in Fig. 5. This time the relative contribution of scattering into Rayleigh waves small. The overall value of the numerical factor in Eq. (30) is also smaller than that in Eq. (23) for the longitudinal wave (for example, at $s = 0.5$ the difference amounts to almost an order of magnitude). This may appear to indicate that a surface with a small correlation length is more specular for transverse than for longitudinal waves. It should be noted, however, that this comparison is made at an equal wavelength. A comparison made at an equal frequency, on the other hand, yields a larger scattering power for the transverse incident wave since its wavelength is smaller and we have to account for a factor $(k_t/k_l)^4$, equal to 16 in the example with s=0.5.

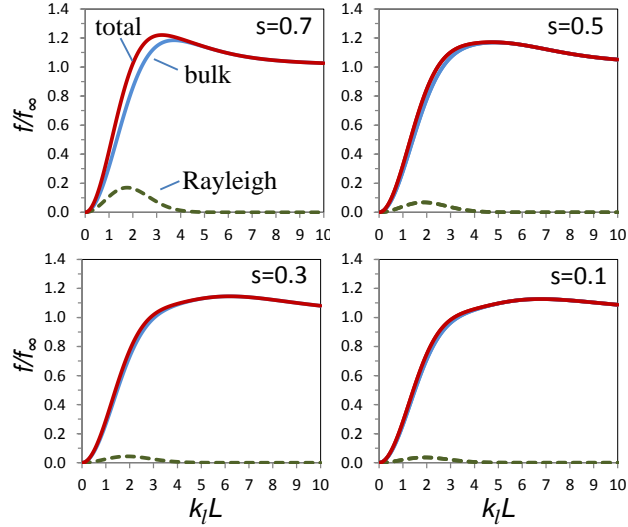

FIG. 6. (Color online) Normalized diffuse scattering probability for the normally incident transverse wave vs. the product of the acoustic wavevector and the correlation length for different values of the velocities ratio $s$. Contributions of scattering into bulk and Rayleigh surface waves are shown as indicated in the upper left panel.

Figure 6 shows general case results for the Gaussian autocorrelation function. The main trends are similar to the case of the incident longitudinal wave; however, the contribution of scattering into the Rayleigh wave is now much smaller and the dependence on $s$ is less pronounced. In particular, no sharp maximum appears at small values of $s$. As discussed above, the latter is due to the fact that scattered transverse and Rayleigh waves have much smaller wavelengths than the incident longitudinal wave. This phenomenon is unique to the case of the incident longitudinal wave and does not arise in the case of the incident transverse wave whose wavelength is always smaller than that of the longitudinal wave and just a bit larger than that of the Rayleigh wave at the same frequency.

## V. OBLIQUE INCIDENCE

It would be straightforward to extend the method described in Sec. III onto the more general case of an obliquely incident longitudinal or transverse wave. However, the calculations become much more tedious because the flat surface solution will generally contain both transverse and longitudinal reflected waves and involve multiple non-zero components of the stress tensor. Even for the simplest case of an horizontally polarized transverse incident wave, where specular reflection is not accompanied by mode conversion, the equation analogous to Eq. (28) will contain multiple terms involving products of real and imaginary components of $\tilde{G}_{11}$, $\tilde{G}_{22}$, $\tilde{G}_{12}$ and a Fourier-component of the autocorrelation function $\tilde{C}(\mathbf{k} - \mathbf{k}_0)$, where $\mathbf{k}_0$ is the in-plane wavevector component of the incident wave. The oblique incidence breaks the symmetry that made it possible to reduce the final result to one-dimensional integrals; in the general case, two-dimensional numerical integration will be necessary.

In order to avoid tedious mathematics, we will limit the discussion of the oblique incidence to the case of a liquid medium. Even though we have seen that beyond the Kirchhoff approximation there are significant differences between the cases of liquid and solid media, this discussion will still be instructive in terms of looking into the effect of the oblique incidence on the general trends discussed in previous sections. We consider an acoustic wave incident on a rough surface of a liquid half space at an angle $\theta$ to the normal and follow the sequence of steps described in Sec. III, which is made easy by the absence of shear stresses in liquid. The result obtained for the scattered power

$$f = \frac{1}{\pi^2} \eta^2 k_l^3 \rho c_l^2 \cos\theta \int \tilde{C}(\mathbf{k} - \hat{\mathbf{j}} k_l \sin\theta) \operatorname{Im}\tilde{G}_{33}(\mathbf{k},\omega) d\mathbf{k} , \qquad (31)$$

where $\hat{\mathbf{j}}$ is the unit vector along *y*, differs from Eq. (14) only by a factor of $\cos\theta$ and by a shift in the argument of the spectral autocorrelation function by the in-plane component of the incident wavevector. Using Green's function from Eq. (26) we get a known result [41],

$$f = \frac{1}{\pi^2}\eta^2 k_l \cos\theta \int_{k<k_l} \tilde{C}(\mathbf{k}-\hat{\mathbf{j}}k_l \sin\theta)\left(k_l^2 - k^2\right)^{1/2} d\mathbf{k}, \qquad (32)$$

The limiting case of a large correlation length is well documented in the literature [10,41]. In this case, $\tilde{C}$ is a narrow function compared to the square root in the integrand. If the latter is changing slowly within $\sim 1/L$ from $\mathbf{k}_0 = \hat{\mathbf{j}}k_l \sin\theta$, then we can then replace *k* in the square root by $k_l \sin\theta$, which leads to a result that

$$f_\infty = 4\eta^2 k_l^2 \cos^2\theta, \qquad (33)$$

which, again, perfectly agrees with Ziman's equation. The condition for the square root to be a slowly varying function within $\sim 1/L$ from $\mathbf{k}_0$ is $k_l L \gg 1/(1-\sin\theta)$, which becomes increasingly stringent for large incidence angles and necessitates a special treatment of grazing incidence [10, 41]. In the opposite limiting case of a small correlation length, $k_l L \ll 1$, the autocorrelation function in the integrand of Eq. (32) can be replaced by its value at *k*=0, yielding

$$f_0 = \frac{2}{3\pi}\cos\theta\eta^2 k_l^4 \tilde{C}(k=0). \qquad (34)$$

This simple result does not appear to have been reported in the literature even though it would be straightforward to obtain it within the framework developed in Refs. [10,11,41]. For the Gaussian correlation function we get

$$f_0 = \frac{2}{3}\eta^2 k_l^4 L^2 \cos\theta. \qquad (35)$$

Thus the observation that in the limit of a small correlation length the scattered power scales as $\eta^2 k_l^4 L^2$ remains valid for oblique incidence. However, the presence of $\cos\theta$ will make scattering vanish at grazing angles, consistent with the intuitive notion that surfaces tend to become specular for grazing incidence.

## VI. DISCUSSION

As we have seen, Ziman's formula is only accurate in the Kirchhoff approximation limit of a large *kL*. However, in this limit there is a caveat pertaining to using Eq. (1) in thermal transport models [5]: if the correlation length is large, the diffusely scattered field will form a narrow forward lobe around the specular direction; as a result, calculations based on the assumption that scattered field is isotropic will overestimate dissipation of the heat flux due to the boundary scattering. On the other hand, modeling interfacial roughness as atomic disorder [26,27], which implies a small correlation length, will typically underestimate boundary scattering. A case in point is the attenuation of sub-THz coherent phonons is GaAs-AlAs superlattices [47]: experimentally measured extrinsic scattering rates (i.e., scattering by interface roughness and defects) were orders of magnitude greater than the atomic disorder model predicted. Incidentally, the experimental scattering rate scaled with frequency as $\omega^{2.7}$, indicating an intermediate case between the limits of $\omega^2$ and $\omega^4$ scaling. It should be noted that scattering of sub-THz phonons by interface roughness in a superlattice is one case where the small perturbation approach would be well justified as losses in a single scattering event are typically small: for example, one can see ~0.3 THz coherent phonon wavepackets cross over 400 interfaces without much loss at 79 K [47]. An rigorous analysis of phonon scattering by interfacial roughness in a superlattice will require a separate treatment as the problem is different from scattering by a free surface of a bulk material; however, general trends are expected to be similar to the ones discussed here.

Another point that has been made clear by our analysis is that one should be very careful with using models for scalar waves (essentially acoustic waves in liquid) [30,48] or borrowing results from optics [49] when analyzing boundary scattering of phonons. It is only in the Kirchhoff approximation that the specularity is the same for waves of any nature. Beyond the Kirchhoff approximation the specularity depends on whether we are dealing with a longitudinal or transverse wave, or a scalar wave in liquid; in particular, we have seen that at small correlation lengths the scalar wave model yields a diffuse scattering probability which is by more than an order of magnitude smaller than for a longitudinal wave in a solid.

A much harder question is what happens beyond the small perturbation approximation. While models going beyond the Born approximation have been developed in the context of thermal conductivity of nanowires [25,30], with analysis conducted in terms of eigenmodes of nanowire waveguides, the issue of the specularity of a rough surface for an incident plane wave beyond the Born approximation remains open. An intriguing issue is the so-called "diffuse mismatch" model of the thermal boundary resistance [50] based on the conjecture that a phonon arriving to a very rough interface forgets which side it came from and gets scattered with probabilities proportional to the densities of states in the materials to either side of the boundary. The diffuse mismatch model is obviously incompatible with the Kirchhoff approximation in which the surface is locally flat and transmission/reflection are determined by the impedance mismatch. Furthermore, this model leads to a seemingly paradoxical result that an acoustic wave incident on a rough solid/air interface from inside the solid will be mostly scattered into the air, with a very small fraction of the incident power scattered back into the solid. Can this behavior be reproduced by any physically realistic model of interface roughness? The author hopes that this report will stimulate interest to this and other interesting problems of wave scattering from rough surfaces arising in the thermal transport context.

## VII. SUMMARY

We have analyzed scattering of normally incident longitudinal and transverse waves by a randomly rough surface of an elastically isotropic solid within the small perturbation approach. For an isotropic autocorrelation function of the surface roughness, the specularity reduction (i.e. the diffuse scattering probability) has been expressed in the form of straightforward one-dimensional integrals. In the limiting case of a large correlation length compared with the acoustic wavelength, the specularity reduction is equal to $4\eta^2 k^2$, in agreement with the known Kirchhoff approximation result given by Ziman's formula, whereas in the opposite limiting case of a small correlation length, the specularity reduction has been found to be proportional to $\eta^2 k^4 L^2$, with the fourth power dependence on frequency as in Rayleigh scattering. It has been found that beyond the Kirchhoff approximation the specularity depends on whether the medium is solid or liquid, and in the former case on whether the incident wave is longitudinal or transverse. In particular, scattering into Rayleigh surface waves has been found to play a large role in the specularity reduction for the longitudinal incident wave. In this case, scattering into transverse and Rayleigh waves results in a distinct maximum of the diffuse scattering probability at an intermediate value of $L$, which becomes increasingly pronounced as the Poisson's ratio of the medium approaches 1/2. It is hoped that the present study will help researchers working in the fields of solid state acoustics, phonon physics, and thermal transport in understanding issues related to specularity of rough surfaces. In particular, the results have indicated that thermal transport models based on Ziman's formula are likely to overestimate the heat flux dissipation due to boundary scattering, whereas modeling interface roughness as atomic disorder is likely to underestimate scattering.

## ACKNOWLEDGMENTS

The author greatly appreciates illuminating discussions with Andreas Mayer. This work was supported as part of the S$^3$TEC Energy Frontier Research Center funded by the U.S. Department of Energy, Office of Science, Basic Energy Sciences under Award DE-SC0001299.

## APPENDIX

This Appendix presents detailed results for the transverse incident wave. For an isotropic half-space, the spectral surface Green's function $\tilde{G}_{11}$ is given by [37,39]

$$\tilde{G}_{11} = \frac{1}{\rho c_t^2}\left[\frac{k_t^2 k_x^2 \left(k^2 - k_t^2\right)^{1/2}}{k^2 R} + \frac{k_y^2}{k^2\left(k^2 - k_t^2\right)^{1/2}}\right] + \frac{i\pi H}{\rho c_t^2}\frac{k_x^2}{k^2}\delta(k - k_R), \tag{A1}$$

where

$$H = \frac{\beta^2}{2}\left(1 - \beta^2\right)^{1/2}\left[\frac{8(2 - \alpha^2 - \beta^2)}{(2 - \beta^2)^2} + \beta^4 - 4\right]^{-1}. \tag{A2}$$

Plugging this Green's function into Eq. (28) and assuming that the autocorrelation function is isotropic, we get the main result,

$$f = \frac{1}{\pi}\eta^2 k_t^4\left[\int_0^s \frac{x(1-x^2)^{1/2}\tilde{C}(xk_t)}{4x^2(1-x^2)^{1/2}(s^2-x^2)^{1/2}+(2x^2-1)^2}dx + \int_s^1 \frac{x(1-x^2)^{1/2}(2x^2-1)^2\tilde{C}(xk_t)}{16x^4(1-x^2)(x^2-s^2)+(2x^2-1)^4}dx\right.$$

$$\left.+\int_0^1 \frac{x\tilde{C}(xk_t)}{(1-x^2)^{1/2}}dx + \frac{\pi}{\beta}H\tilde{C}(k_R)\right]. \tag{A3}$$

By setting the spectral autocorrelation function $\tilde{C}$ in the above equation to its value at the zero argument, we find expressions for dimensionless parameters $J_{bulk}$ and $J_R$ in Eq. (30),

$$J_{bulk} = \int_0^s \frac{x(1-x^2)^{1/2}}{4x^2(1-x^2)^{1/2}(s^2-x^2)^{1/2}+(2x^2-1)^2}dx + \int_s^1 \frac{x(1-x^2)^{1/2}(2x^2-1)^2}{16x^4(1-x^2)(x^2-s^2)+(2x^2-1)^4}dx + 1,$$

(A4)

$$J_R = \frac{\pi H}{\beta}.$$